\documentclass[fleqn,usenatbib]{mnras}

\usepackage{newtxtext,newtxmath,xcolor}

\usepackage[T1]{fontenc}

\DeclareRobustCommand{\VAN}[3]{#2}
\let\VANthebibliography\thebibliography
\def\thebibliography{\DeclareRobustCommand{\VAN}[3]{##3}\VANthebibliography}

\usepackage{graphicx}	
\usepackage{amsmath}	

\title[WISE guys not Dyson spheres]{Did WISE detect Dyson Spheres/Structures around {\it Gaia}--2MASS-selected stars?}

\author[Andrew W. Blain]{
Andrew W. Blain,$^{1}$\thanks{E-mail: ab520@leicester.ac.uk}
\\
$^{1}$Physics \& Astronomy, University of Leicester, University Road, Leicester, LE1 7RH, UK\\
}

\date{Accepted XXX. Received YYY; in original form ZZZ}

\pubyear{\the\year{}}

\begin{document}
\label{firstpage}
\pagerange{\pageref{firstpage}--\pageref{lastpage}}
\maketitle

\begin{abstract}
Soon after the release of the WISE all-sky catalogue of $5 \times 10^8$ mid-infrared (IR) objects,  suggestions were made that it could be used to search for extrasolar devices constructed by an advanced civilization to convert a significant fraction of their host star's luminosity into useful work: "technostructures", "megastructures" or "Dyson spheres/structures", hereafter DSMs, whose inevitable waste heat would be seen by WISE at mid-IR wavelengths. However, a trawl of several million potentially-habitable {\it Gaia}-detected stars for mid-IR-excess signatures is fraught with danger, due to both noise from such a large sample and, more importantly, confusion with the emission from dusty background galaxies. In light of a recent claim of seven potential DSMs in MNRAS, a brief rebuttal appeared on arXiv. Further to this response, the relevance of WISE-detected galaxies is discussed in more detail, leading to a seemingly tight limit on the number and lifetime of DSMs, and indeed intelligent worlds, in the $\simeq 600$-pc-radius region patrolled by {\it Gaia}. However, the detectability of DSMs is questioned: a DSM might extinguish its star at optical/near-IR wavelengths, and thus either not appear or appear anomalously faint in a stellar catalogue. Moreover, a civilization advanced enough to construct a DSM is likely to be advanced enough to use countermeasures to mask its presence from us.
\end{abstract}




\section{Introduction}

First suggested in science fiction \citep{SciFi}, \citet{Dyson} discussed hypothetical large-scale circumstellar structures built to intercept starlight and power an advanced civilization.\footnote{Searching for Freeman-Dyson-related science is difficult owing to his huge volume of work, and the even larger number of citations to Dyson--Schwinger quantum-mechanical methods; see the obituary by \citet{DasGupta}} 
The natural consequence of this interception would be a heat signal exhausted at planetary temperatures. These objects are often described loosely as "Dyson spheres", which might be annular \citep{ForganElvis} or swarms of orbiting objects. Hereafter we describe them as Dyson spheres/megastructured (DSMs).\footnote{
An appropriate acronym? It is already well known in psychiatry as The Diagnostic and Statistical Manual of Mental Disorders.} 
A search for this signature in the catalogue of $\simeq 2.5 \times 10^5$ sources detected by {\it IRAS} \citep{SHN} was described by \citet{Carr}: they considered how many {\it IRAS} sources could be waste heat from such structures. While this was an interesting exercise, the number of objects searched was modest, and so \citet{Carr} did not yield a very significant result. Post-{\it IRAS}, both {\it MSX} \citep{Stair} and {\it Akari} \citep{Mura} provided substantial but partial sky coverage; they probed deeper than IRAS, but did not qualitatively change the situation regarding searches for DSMs.

A major step forward was made by WISE \citep{Wright2010}, yielding a sensitive all-sky mid-infrared (IR) survey with $\sim 10^9$ catalogued sources, well over a thousand times more than {\it IRAS}, and reaching orders of magnitude deeper than {\it IRAS} at the common mid-IR wavelengths of 12-25\,$\mu$m. 

Carrigan's approach has been extended to the WISE catalogue \citep{Cutri} this year \citep{CH, Suazo}. The combination of the WISE catalogue with precise locations and optical photometry for about five million long-lived main-sequence stars from the {\it Gaia} catalogue \citep{Gaia},\footnote{https://www.cosmos.esa.int/web/gaia/dr3} 
and near-IR photometry from the near-IR all-sky 2MASS survey \citep{2MASS}\footnote{https://irsa.ipac.caltech.edu/Missions/2mass.html} 
put this search on a more secure statistical footing. 

Seven candidate DSMs were suggested by \citet{Suazo} from a WISE search for five million M stars in the {\it Gaia} catalogue. At face value, this seemed to be perilous, given the large areal density of WISE-detected galaxies with substantial excess 12- and 23-$\mu$m emission and much less luminosity at wavelengths shorter than 5\,$\mu$m.\footnote{
While Facebook posts are not a legitimate source, they do provide a record to friends of my incredulity at the claim
https://www.facebook.com/andrewwblain/posts/pfbid0ifrq4qing1S6E9DVnv gbdNU14xaBP6kP71dFikEwyF3oy4MNx2TkyVbeXXuwLxLFl. (Zuckerberg \& Blain 2024)} 
A similar analysis by \citet{CH} claimed less sensational results, and stated that WISE mid-IR excesses were found associated with long-lived stars, but with plausible astrophysical explanations. 

Soon after \citet{Suazo} appeared, \citet{RGS} cast doubt upon the claims for the three cases with deep radio survey data available. They identified faint accurately-located radio sources at positions offset by 0.5, 4.9 and 5.0\,arcsec from the precise {\it Gaia} star positions, with IR-to-radio flux ratios consistent with the presence of a background ultraluminous WISE-detected galaxy, a so-called "HotDOG" at redshift $z \simeq 1-3$ \citep{ChaoWei,DS,DS2}. The presence of this information in the literature motivated this paper, and these radio offsets assist the analysis significantly. 

\section{Dyson spheres/megastructures}

The idea that civilizations would run out of resources on a planetary scale led to the natural suggestion that they could devise means to harvest energy from their host main-sequence star \citep{Dyson,Kardashev}; or, more exotically, from a White Dwarf \citep{SO}, a pulsar \citep{Oz17}, or even from a blackhole \citep{Hsaio}, in which the cosmic microwave background (CMB) would be absorbed, and waste heat exhausted directly into the blachhole. Tapping non-main-sequence sources would presumably require the civilization first to master interstellar travel to keep their structures safe from stellar winds or explosions at the end of core burning. The detectability of such a DSM would derive from its dimming of the stellar object being harvested, and from its exhaust heat, the relevant fraction of the star's luminosity, and likely to appear at IR wavelengths, similar in spectral energy distribution (SED) to debris disks, asteroid belts and planets. 

Summaries of the developed ideas for DSMs have been reviewed by \citet{Wright2020}. Searches for such objects have ranged from the more physical \citep{Carr, CH, Zuck}, to the more speculative \citet{Suazo}, and all the way to the seemingly outlandish \citep{Loeb}. 
Suggestions of even larger constructed objects tapping the energy from whole galaxies have also been made \citep{GWM,L}.

\subsection{Waste heat}

Waste heat from a DSM is unavoidable: stellar lifetimes are huge as compared with energy processing times in a DSM, and so thermal balance is essential.  

\subsection{Thermal balance}
\label{TB}

It is possible that the geometry of a DSM could be complex, and multi-layered. However, for simplicity a spherical object is assumed to discuss temperatures, masses and dynamical times. 

Assuming all, or at least a traditional cosmologist's "significant fraction" of the luminosity of a star, $L$, is intercepted by a DSM with radius $R$, the spherical shell would reach a temperate $T$, assuming blackbody absorption: 
\begin{equation} 
\left[\frac{L}{{\rm L}_\odot} \right] \simeq \left[\frac{R}{1\,{\rm AU}} \right]^{2} \left[\frac{T}{300\,{\rm K}}\right]^{4}.
\label{LRT}
\end{equation}
As a result, given the strong dependence of $L$ on stellar mass $M$, long-lived low-mass stars, as found in great abundance in the Milky Way, might be harvested using a relatively small DSM by Solar-system standards. Scaling mass to luminosity as $L \propto M^{3.6}$, the temperature of a DSM $T \propto M^{0.9} R^{-1/2}$. At constant temperature $R$ scales with stellar mass as $M^{1.8}$. 

It seems reasonable that the elements of a DSM must be cooler than about 1000\,K to ensure structural soundness. At the lower temperature limit, if heated only to a 3-K microwave-background temperature, or an ambient interstellar medium (ISM) temperature of order 10-20\,K, there would be no point being near a star, and no way to exhaust heat. For the Solar system, temperatures of 1000, 20 and 3\,K correspond to $R \simeq$ 0.09, 225 and $10^4$\,AU respectively. $10^4$\,AU is 0.05\,pc, and so even this huge structure is still  closer to its host star than to its neighbours. 

The mass of a DSM with thickness $t$ would be approximately, 
\begin{equation}
\left[ \frac{M}{{\rm M}_\odot} \right] = 
1.5 \times 10^{-7} \left[ \frac{R}{{1\,{\rm AU}}} \right]^2 \left[\frac{\rho}{1000\,{\rm kg}\,{\rm m}^{-2}} \right] \left[ \frac{t_1}{1\,{\rm mm}} \right].
\label{MR}
\end{equation}
It is difficult to see $t$ could be thinner than a few millimetres, in order to collect and channel power at the kW\,m$^{-2}$ level, following the active layer of terrestrial and space-borne solar panels. Regardless of technology, I think it is inevitable that an absorber must be at least a few wavelengths thick, making 10\,$\mu$m an absolute limit.
A 1-mm thick 1000-kg-m$^{-3}$ dense spherical shell at 1\,AU would have a mass of $3 \times 10^{23}$\,kg. This is about 0.015\,per cent of the mass of Jupiter, 5\,per cent of the mass of Earth, and about 150 times the mass of the main asteroid belt \citep{Petit}. Using Jupiter's material - assuming very optimistically that Jupiter is all metals/suitable construction material, and accessible, a 10-m-thick shell could be built at Earth's orbit, and a 0.4-m-thick shell at Jupiter's 5\,AU. The use of resources on a Solar-system scale is discussed by \citep{ForganElvis}, as a probe of extraterrestrial intelligence. 

Building a 1-mm-thick DSM from the Sun's asteroid belt material would require $R \le 0.09$\,AU, reaching $T=1000$\,K. Using the whole of the mass of the Earth instead would correspond to $R \le 4.5$ and $T \ge 140$\,K; the corresponding values for the mass of Jupiter are $R \le 81.3$\,AU and $T \ge 33$\,K. Assuming that 1\,mm is the absolute minimum thickness, these radii would be strict upper limits. 

Jupiter-mass planets appear to be common in our neighbourhood, and so an AU-scale DSM seems possible for planetary systems.  If a Jupiter-mass DSM were at the absolute maximum outer radius $R=10^4$\,AU as discussed above based on temperature, the thickness would be an impractically thin 4\,nm. A DSM must thus be very much smaller than the typical interstellar distance, and likely smaller than 100\,AU.

Simple isotropic energy balance is not necessarily a direct guide to detectability, however: if a civilization is able to build an AU-scale DSM, it seems reasonable that it could take deliberate countermeasures to hide its presence. Ways to evade detection could be to beam mimicked stellar signatures at potentially observant stars in its neighbourhood, and to directional exhaust waste heat away from them. 

\section{WISE-detected background galaxies} 

WISE catalogued "the most luminous galaxy" \citep{DS, DS2}. There are of order 2200 similar examples, labelled HotDOGs \citep{ChaoWei}; also see Eisenhardt et al. (in prep.). HotDOGs have luminosities of order $10^{14}$\,L$_\odot$, known examples lie out  of the Galactic plane, more than an arcsecond away from stars, and at cosmological distances. HotDOGs have both the appropriate SED and flux density to be mistaken for a 1-L$_\odot$--100-K DSM at a distance of order 100\,pc. Less-luminous IR galaxies are too faint/rare to mimic a DSM: a $10^{12}$-L$_\odot$ IR galaxy (ULIRG) at a distance of 100\,Mpc is also comparable in brightness to a DSM candidate; however, the density of ULIRGs is only $10^{-7}$\,Mpc$^{-3}$ at low redshifts, meaning there are only a handful over the sky. 

The RMS positional uncertainty for a WISE source detected at modest signal-to-noise ratio in the two longest-wavelength bands W3 and W4 at 12 and 23\,$\mu$m is of order 2\,arcsec, while {\it Gaia}-measured stars are located to milliarcsec precision. The likelihood of matching a WISE and {\it Gaia} source is thus determined by the WISE positional uncertainty. 
The isotropic surface density of HotDOGs over the whole sky is about 0.1\,deg$^{-2}$.\footnote{\citet{RGS} quote 0.032\,deg$^{-2} = 9 \times 10^{-6}$\,arcsec$^{-2}$ (sic); however, the full HotDOG catalogue is a little larger, with 2220 found over 70\,per cent of the sky, yielding 0.1\,deg$^{-2} = 7.7 \times 10^{-9}$\,arcsec$^{-2}$.} 
The area within an angle $\theta"$ of a {\it Gaia}-mapped star in arcseconds is $\pi \theta"^2$\,arcsec$^2 = 2.4 \theta"^2 \times 10^{-7}$\,deg$^2$, and so a survey this close to five million {\it Gaia} stars would search a total area of 1.2$\theta"^2$\,deg$^2$. If $\theta"=1$\,arcsec, \citet{Suazo}'s listed search radius, then the likelihood of a match between an M star and a HotDOG is about 12\,per cent: $5 \times 10^6 \>\times\> 2.4 \times 10^{-7}$\,deg$^2 \>\times\> 0.1$\,deg$^{-2} \simeq 0.12$. However, \citet{RGS} found their three radio sources at radio--{\it Gaia} separations of 0.4, 4.9 and 5.0\,arcsec. Radio positions are accurate to fractions of an arcsec and {\it Gaia} positions are precise; however, WISE the positions are accurate only to about 2\,arcsec. The radio-{\it Gaia} positional offsets thus exceed the 1-$\sigma$ uncertainties in the WISE positions, and so the effective search radius from \citet{Suazo} was actually larger: likely $\theta" \simeq 5-6$\,arcsec: searching $5 \times 10^6$ stars, it is reasonable to sample a radius many times larger than the RMS positional uncertainty in the catalogue. The 12\,per cent chance of a 1-arcsec-radius match is then increased by a factor $\theta"^2$, and thus a handful of {\it Gaia}--WISE matches would be expected, as found: pure abundance matching in the WISE catalogue can account for all the candidates. The candidate DSMs \citet{Suazo} are consistent with chance alignments of background galaxies with {\it Gaia} stars.

\section{DSM Statistics} 
\label{DSMSTAT}

There is a probability $P \leq 2 \times 10^{-7}$ that a particular surveyed star has a DSM around it; subject to ignoring the effects of any countermeasures its occupants were taking to evade detection (see Section\,\ref{AC}). 

Stellar lifetimes impose no constraints at these stellar masses, and so this probability can be converted directly into an estimate of the chance that hypothetical planetary-system-modifying, DSM-building civilizations develop amongst {\it Gaia} stars, and of their lifetime or at least the lifetime of their Gyr-lingering post-apocalyptic wreckage \citep{Loeb}. The fractional abundance of DSMs among {\it Gaia} stars today $P$ is the product of the probability that a star ever develops a DSM $F$, and a typical DSM lifetime $t$ in years: $P \simeq F [t/10\,{\rm Gyr}] \leq 2 \times 10^{-7}$, and so $Ft \leq 2000$\,years. If a DSM was built in every stellar system, $F=1$, then $t < 2000$\,years. Two thousand years is likely much too short to build a DSM \citep{ElvisMill}, and definitely much less time than required for its decay after being abandoned \citep{Loeb}. DSMs are thus definitely not ubiquitous in the history of {\it Gaia} stellar systems. For an assumed DSM lifetime $t=1$\,Gyr, $F < 2 \times 10^{-6}$, and so long-lived DSMs are very rare. 

For long-lived DSMs, we can use a Drake equation to include the probability that intelligent life, having appeared in a stellar system, develops a DSM $P_{\rm DSM}$. The usual probability that intelligent life evolves is $P_{\rm Drake}$, and so $F = P_{\rm Drake} P_{\rm DSM}$. We know $F \le 2000 t^{-1}$, and that $P_{\rm Drake}$ is probably small. $P_{\rm Drake} P_{\rm DMS} < 2000\,{\rm yr} \> t^{-1}$, and assuming a reasonable lifetime $t = 1$\,Gyr, $P_{\rm Drake} P_{\rm DMS} \le 2 \times 10^{-6}$. The maximum lifetime for a DSM to date $t \simeq 10$\,Gyr, imposing a more stringent $P_{\rm Drake} P_{\rm DMS} \le 2 \times 10^{-7}$. Both of these $P$ quantities are interesting, perhaps $P_{\rm Drake}$ the most, and neither can exceed unity. If it is inevitable that intelligent life eventually builds a DSM, then $P_{\rm DSM}=1$, and so $P_{\rm Drake}<2\times10^{-6}$: intelligent life is very rare. If intelligent life is ubiquitous, then $P_{\rm Drake}=1$, $P_{\rm DSM}<2\times10^{-6}$, and so building DSMs is very unlikely. Various more complex joint constraints on $t$ and $P$ values or their distributions can be considered. 

It therefore seems likely that, even without countermeasures being taken to hide a DSM, there are zero DSMs within the extensive {\it Gaia} survey volume, of order 300\,pc around us, a few percent of the volume of the Mily Way disk. An IR follow-on mission to {\it Gaia} would cement this result more clearly, penetrating further through the ISM. 

\section{Alien countermeasures} 
\label{AC}

A civilization might be wise not to attract interstellar attention while harvesting its star's power.
\footnote{{This has been labelled a "dark forest" picture after the title of a 2015 novel by \citet{Cixin}, but has not been granted mainstream scientific kudos by association with rigorous investigations, like DSMs \citep{Dyson, Wright2020}.}}

It seems reasonable that a civilization capable of building AU-sized DSMs could beam confusing/masking/spoofing multiwavelength signals towards nearby stars to appear as ordinary stars rather than DSMs. It could also take advantage of exhausting waste heat anisotropically, to reduce the IR signature towards local stars, reducing the WISE flux sought by \citet{CH} and \citet{Suazo}. 

Stars fill a tiny fraction of the sky, and even if civilizations could send probes throughout their solar systems, the distances from star to the furthest probe are still likely to be modest compared with the distance between stars. DSMs themselves are likely to be a maximum of order 100s of AU in size. 

If aliens using a DSM decided to beam an artificial stellar signal to the regions around neighbouring stars where observers may lurk, either on planets or any launched probes, it would not be a huge challenge either energetically or to their technological skills. Spoofing a nearest neighbour stellar system that nothing suspicious was occurring would require a fake starlight signal extending out to a distance of order 100\,AU  to cover all possible planets and probes; this corresponds to a beam that is only a fraction $2.5 \times 10^{-8}$ of the sky at 1\,pc. As the distance to stars towards which countermeasures are required $D$ increases, the solid angle that must be illuminated to hoodwink all the planets and probes around each star is reduced, scaling with $D^{-2}$. The power that must be broadcast per solid angle to mimic the flux of a star within this countermeasure beam of course remains constant, and so the total power required per spoofed star scales as $D^{-2}$. However, the number of stars that require a countermeasure beam in their direction increases as $D^3$ until the edge of the disk is reached. The total power required thus increases as $D$, out to $D \simeq 100$\,pc. On larger scales, the edge of the Galactic disk is reached, and the number of stars needing countermeasure signals increases more slowly with increasing distance $D$, tending eventually to increase with $D^2$. The total power that must be broadcast thus increases more slowly than $D$ for $D > 100$\,AU. The total power required for targets at all distances is thus only a few hundred times that required to spoof a single nearest neighbour at 1\,pc. On larger kiloparsec scales potentially suspicious stars are cloaked by the ISM if they are in the disk, and stars are rare out of it. The total energy demand for countermeasure signals is thus equivalent to only about a few millionths of the stellar luminosity. 

Heat engines could modify the blackbody equilibrium condition to cool the bulk of a DSM, with a higher-temperature radiative exhaust directed out of the Galactic plane/away from potential viewers. 
A heat engine could potentially exhaust at a very high temperature, matching stellar colours, which would make the DSM appear as an anomalously bright star from some directions. 
To avoid any gradual jet thrust of the cloaked star's exhaust raising suspicions, changes in the emission directions, or bipolar exhausts could be used. Over 1\,Gyr, the thrust from emitting 1\,${\rm L}_\odot$ from a 1-M$_\odot$ star in a constant direction would lead to a speed of 20\,km\,s$^{-1}$. A more massive star would have a greater thrust-to-mass ratio, and thus gain speed more rapidly. The speed depends on stellar mass as $\Delta v \propto M^{2.6}$. For 1.5(0.7)\,M$_\odot$ stars, after 1\,Gyr 
$\Delta v = 57(7.9)$\,km\,s$^{-1}$. The effect is thus negligible for a subsolar star, but a long-term harvester of the energy from a supersolar star would need to act to avoid their star appearing unusual to an alien {\it Gaia}.

A civilization capable of building a DSM should also be able to keep a careful watch on all its neighbours for any signs of technological ability or  observing suspiciously. An aperture synthesis array with a 10/100-AU baseline can resolve detail of order 100/10\,km in size at 100\,pc. Nearest neighbours at distances of a few parcsecs could thus potentially be examined passively for radio emissions on sub-km scales. Active radar would require high powers, but careful signal design and modulation might allow it. Operating such a device at CMB wavelengths of order 10\,mm makes sense to reduce the chance of detection. Gravitational lensing might offer a route to obtaining sufficient power \citep{Tusay}.
Radio and far-IR observations would not be immune to spoofing. An aperture synthesis telescope could be jammed/mislead by emitting carefully-phased and directed radio signals. 

Curious aliens could always see the home of DSM builders as a regular unexciting G/M star, with an appropriate level of variable circumstellar opacity and magnetospheric activity.\footnote{If desired, a DSM owner could even fake a background HotDOG.} 

\section{Tests and checks} 

Various features of the distribution and properties of {\it Gaia}-2MASS--WISE matches to potential DSMs could be used to test their reality. 

\begin{enumerate} 

\item {\bf Changing the {\it Gaia}--WISE search radius.} An increase in the {\it Gaia}--WISE catalogue search radius for coincident sources $\delta$ should lead to a direct increase in the number of DSM candidates, scaling as $\delta^2$ for precise positions. WISE positional uncertainties would probably reduce the scaling factor, in a way that could be simulated accurately. This is probably the least expensive test of the DSM hypothesis. Searching for WISE objects with the same mid-IR properties at random rather than {\it Gaia} star positions should give a similar abundance of matches if background galaxies are randomly located. Astrophysical mid-IR emission from stars would still appear as a correlated signal \citep{CH}.

\item {\bf Deep follow-up radio/submm imaging.} \citet{RGS} suggested deeper radio imaging for all seven DSM candidates from \citet{Suazo}.\footnote{"When your tool is a hammer, every problem tends to look like a nail?"} Finding radio emission in all candidates at a position offset significantly from the {\it Gaia} star, and with a flux at least characteristic of a radio-quiet WISE HotDOG, would make a projected background HotDOG a much more likely explanation for the WISE-{\it Gaia} sources than a DSM. 

Radio emission from a distant background galaxy should be extended on scales of at least 0.1-1\,kpc, corresponding to an angle of order 0.1\,arcsec. This is a much larger than the 0.1-milliarcsec subtended by an active star at 100\,pc, and matches the size of a 10-AU-scale DSM broadcasting radio countermeasure signals from the same distance. Interplanetary-scale-baseline radio imaging could perhaps reveal the distribution of any spoofing antennas. An interstellar electronic-intelligence arms race could develop to hide and seek such emitters. 

Radio-recombination-line spectra could perhaps reveal a redshift from a line in a distant galaxy responsible for a mid-IR excess in very deep radio observations with the Square Kilometre Array (SKA), confirming the background galaxy explanation beyond doubt. 

A more fruitful approach might be to make deep ALMA (sub)mm-wave images of the 10-arcsec-wide region surrounding the {\it Gaia} star for the DSM candidates. A resolved detection that is either offset from the star, or coincident but with the morphology of a distant galaxy, would confirm a background galaxy origin for the WISE emission, and simultaneously extend the current sample of tens of HotDOGs with high-resolution (sub)mm-wave imaging. There is a $\sim 10$\,per cent chance of finding a spectral line and thus a redshift in such an observation. Furthermore, the bright star in the immediate vicinity would prove useful for carrying out ground-based near-IR adaptive optics (AO) imaging.

\item {\bf Resolved near/mid-IR imaging}. Deep ground-based AO or {\it JWST} imaging could also resolve emission characteristic of the SED and subarcsecond morphology of a distant galaxy \citep{DS}, in contrast with the expected point source at the exact {\it Gaia} stellar position for a DSM. This would definitively rule out a DSM. At 2\,$\mu$m, a 30-m telescope has a diffraction limit of about 10\,milliarcsec.  

\item {\bf The distribution of DSM candidates on the sky}. DSMs would be expected to be distributed across the sky like Galactic stars, in contrast with the expected isotropic distribution of distant HotDOGs, independent of Galactic latitude $b$. The distribution of DSM candidates should allow a statistical view of whether they shadow the distribution of {\it Gaia} stars or not.\footnote{Of course, if advanced civilizations decided to plan a Gyr-duration joyride out of the Galactic plane on a waste-heat jet, this could modify the distribution.} The galactic latitude of the current seven candidates is very non-uniform, all but one lying in the range $-31 > b > -57$\,deg; they all lie between 143-275\,pc away. 

\item {\bf Measuring parallaxes and proper motions}. Proper motion of mid-IR DSM candidates in the shorter WISE bands are potentially available from the NeoWISE programme \citep{Mainzer} over its 2014-2024 time baseline, for comparison with the very accurate {\it Gaia} parallax and proper motion data for the stars. The WISE and {\it Gaia} objects must not separate for a DSM, while of course distant HotDOGs are fixed on the sky. NeoWISE recently came to an end, but the forthcoming NASA NEO Surveyor (NEOS) is due to continue complementary coverage of much of the sky in 2027, going deeper and with better spatial resolution \citep{NEOS}.\footnote{https://neos.arizona.edu}

\item {\bf Interstellar travellers}. If \citet{Loeb} is correct, then the idea of DSMs in the Milky Way can be tested by watching to see where interstellar objects are coming from. Detailed knowledge of the Galactic potential from {\it Gaia} could then lead us directly to any cosmic vandals launching intentional probes or to Ozymandian DSMs decaying from asteroid collisions, although it seems implausible that many collisions taking place at interplanetary speeds would pitch much of the debris from a DSM onto an interstellar path. There is no problem with timescales of interstellar travel: in 10\,Gyr an object cruising between stars at 30\,km\,s$^{-1}$, matching the observed speed of 'Oumuamua \citep{Meech, Elvis22}, could cover trans-Galactic distances $\sim 30$\,kpc. 

\end{enumerate} 

The results of follow-up high-resolution imaging would have the additional benefit of swelling the catalogue of HotDOGs with excellent multiwavelength data \citep{ChaoWei}. Inevitably, the current HotDOG catalogue misses examples in the small fraction of sky very close to Galactic stars.  

\section{Additional comments} 

\subsection{The physical extent of DSMs and microlensing} 

In Section \,\ref{AC} the total area of sky requiring countermeasure broadcasts to mask a DSM from observations from other stars and their associated planets and probes was discussed. A G/M star and its planetary system with radius $R$ at a distance $D$ subtends an angle $2R/D$; as $D$ exceeds a parsec, this angle cannot be much larger than a few tens of arcsec for an immediate neighbour with $R \simeq 100$\,AU, while the star itself is four orders of magnitude smaller, $\sim 2$\,milliarcsec. At 500\,pc, these angles are two orders of magnitude smaller, $\sim 0.2$\,arcsec for a planetary system, and a $\sim 10$\,$\mu$arcsec for the star. The angular scale of a DSM would be of order a hundred times larger than the star it encloses, and tens of times smaller than a planetary system: $\sim 100$\,$\mu$arcsec at 500\,pc away.
These angles are all small enough to be subject to microlensing by compact objects that wander past the line-of-sight. 

The number of potential DSMs increases quickly with distance: as $D^3$ out to $\sim 100$\,pc, and then continues to grow more slowly, to approach a $D^2$ dependence on kpc scales. DSMs are thus much more likely to have milliarcsec sizes than anything larger. Furthermore, the optical depth to microlensing increases with distance.

Microlensing is rare, with $\sim 10^{-7}$ observed events per year per star \citet{Mroz} towards the LMC, but poses major problems for hiding DSMs. While unlikely for any star, if the DSM owner is spoofing a few million stars, then a microlensing event will occur every few years. Microlensing increases the flux density from an object equally at all wavelengths, and conserves surface brightness. Significant magnification occurs if the deflection angle by the lens $\alpha$ (where $\alpha \sim 1.7$\,arcsec for a solar mass) exceeds the offset from the line between observer and source. As the lens moves across the sky relative to the source the magnification rises and falls. The source angular size $\theta$ imposes a limit to maximum magnification. In close alignment, magnification depends on $1/ \alpha$, is limited to $\theta/\alpha$, and can be many hundreds of times for a milliarcsec-scale source. Different regions of a source can be magnified by different amounts, depending on their relative alignment, and larger sources rise and fall in magnification more slowly than smaller ones. Microlensing could thus reveal internal structure of both a DSM or the broadcasting system of a spoofing signal. The most dramatic effect would be for a non-thermal source of stellar light. A laser system would be much more intense than the surface of a star, and thus be subject to more dramatic magnification. It could even potentially scintillate or be occulted by asteroid-sized objects, see \citet{Schli}. 

The DSM owner could seek to preempt microlensing events by watching stars of concern for signs of a reciprocal magnification effect, and adjust their spoofing signal, However, there is a crucial problem: microlensing events typically last for months, and yet the light travel time to stars is longer: years to centuries. Synchronising a spoofing signal would thus be impossible. Even if a microlensing event can be predicted months in advance, the broadcasted spoofing light signal from the DSM intended to confuse an observer on a potentially curious star is already many years away enroute: there is no way to react in time to avoid the curious star seeing the microlensing effect when it arrives. 

A star could also potentially be used as a "solar gravitational lens" telescope, effectively exploiting microlensing on demand. Detectors must be placed several hundred AU away from the star, and directly aligned with the source being observed \citep{Toth}. This effect could be used to closely inspect neighbouring stars and planetary systems, or to boost signals to facilitate potential interstellar communication \citep{Tusay}. If a curious neighbour were to exploit this approach, a DSM would be easier to detect and harder to hide at a wide range if distances: in fact, it could be very hard to evade detection as a DSM if an observer was determined to investigate (Blain, in prep.). 

\subsection{Support for searches} 

The keyword sociology was added because support for WISE-based DSM studies has been provided by the Templeton Foundation,\footnote{https://www.templeton.org} with an unconventional mission statement. Soon after the WISE catalogue was published\citep{Wright2010, Cutri}, this foundation funded a project to search for DSMs, producing acknowledgements in three publications \citep{Wright2014a,Wright2014b,GWM}, and news coverage.\footnote{
https://www.smithsonianmag.com/air-space-magazine/dyson-spheres-still-missing-maybe-impossible-180949538/, 
https://www.dailymail.co.uk/sciencetech/article-3271546/Have-researchers-alien-MEGASTRUCTURE-Researchers-reveal-bizarre-star-say-huge-unknown-object-blocking-light.html} 
\citet{Suazo}'s Project Hephaistos,\footnote{
https://www.astro.uu.se/$\sim$ez/hephaistos/hephaistos.html
} acknowledges the Magnus Bernvall Foundation, named for a Stockholm venture capitalist,\footnote{
https://www.scalexgp.se/medarbetare/magnus-burvall/} 
without an obvious English-language web presence. This foundation is also acknowledged in an earlier SETI-related paper by \citet{EZ15}. 

It is interesting to speculate whether these efforts would have been supported by a more mainstream funding agency; although the Templeton Foundation certainly funds mainstream scientists and studies: for example Frank Drake's grant 44222 and David Charbonneau's grants 43769 and 60862.\footnote{https://www.templeton.org/grant/multiband-and-multidimensional-searches-for-electromagnetic-indicators-of-extraterrestrial-technology, https://www.templeton.org/grant/the-alien-earths-initiative, https://www.templeton.org/grant/opportunity-m-the-fast-track-to-find-other-inhabited-worlds} It has also awarded prizes to individual distinguished astrophysicists: for example Barrow in 2006 and Rees in 2011. 

\section{Conclusions}

\begin{enumerate} 

\item The evidence for signs of star-harvesting structures (DSMs) built by sophisticated extraterrestrial civilizations from a cross-matching study between {\it Gaia}--2MASS-selected stars and the WISE catalogue is weak. Counting possible candidates gives a null result, owing to the surface density of very luminous distant dusty galaxies that share mid-IR characteristics. 

\item The fraction of M stars that leak starlight through surrounding DSMs is thus fewer than 1 in 5 million. If they arise in all stellar systems, then the typical lifetime of a DSM must be less than about 2000\,years. 

\item The product of the fraction of stars that host an advanced civilization and the fraction of advanced civilizations that build a DSM is $s \times 10^{-7}/t$, where $t$ is the typical lifetime of a DSM. 

\item It is reasonable that a DSM builder will be advanced enough to disguise its presence from its neighbours using electromagnetic countermeasures, I thus claim that any estimates of the abundance of DSMs, and derived abundances of intelligent life are formerly lower limits. 

\item Microlensing is a phenomenon that is difficult for a DSM-building civilization to counter and avoid detection. The ultimate SETI approach might be to use the Sun as a gravitational lens to survey stars for indications of any technosignatures, even if the examined aliens are trying actively to hide from view. Follow-up of candidate DSMs by a hypothetical solar gravitational telescope would also make it difficult for them to evade detection.
\end{enumerate} 

\section*{Acknowledgements}

I thank Martin Elvis, first for pointing out that a good limit to the lack of plausible DSM candidates imposes a more severe and interesting limit to their existence in the Galaxy, and second for a very thorough reading of the manuscript and making many excellent suggestions. Remaining errors are entirely my own. I would also like to thank Peter Eisenhardt, Roberto Assef and Jean Turner for helpful comments, and for Tom Jarrett's helpful suggestion at the beginning of June 2024. Ad astra, Tom; rest in peace.

My institution insists that his text must now be added to papers' acknowledgements: "For the purpose of open access, the author has applied a Creative Commons Attribution (CC BY) licence to the Author Accepted Manuscript version arising from this submission".

\section*{Data Availability}

 \citet{CH}, \citet{Suazo} and \citet{RGS} contain discussion and sources of relevant data.



\bibliographystyle{mnras}








\bsp	
\label{lastpage}
\end{document}